\documentclass[10pt,preprintnumbers,amsmath,amsfonts,amssymb,floatfix,twocolumn,prl]{revtex4-2}
\usepackage[utf8]{inputenc}
\usepackage{physics}
\usepackage{breqn}
\usepackage{mathrsfs}
\usepackage{graphics}
\usepackage{xspace}
\usepackage[T1]{fontenc}
\usepackage{graphicx}% Include figure files
\usepackage{dcolumn}% Align table columns on decimal point
\usepackage{bm}% bold math
\usepackage{float}%
\usepackage{color}
\usepackage[capitalise]{cleveref}
\usepackage{chemformula}

\begin{document}
\title{Multiple Coulomb phases with temperature tunable ice rules in pyrochlore spin crossover materials}
\author{Jace Cruddas}
\email{j.cruddas@uq.edu.au}
\affiliation{School of Mathematics and Physics$,$ The University of Queensland$,$ QLD 4072$,$ Australia}
\author{B. J. Powell}
\email{powell@physics.uq.edu.au}
\affiliation{School of Mathematics and Physics$,$ The University of Queensland$,$ QLD 4072$,$ Australia}

\begin{abstract} %must be <600 characters for PRL
Spin crossover molecules have two accessible states: high spin (HS) and  low spin (LS).
We show that, on the pyrochlore lattice, elastic interactions between SCO molecules can give rise to three spin-state ice phases. Each is a ``Coulomb phase'' where a local ice-rule can be mapped to a divergence free gauge field and the low energy excitations carry a spin fractionalized midway between the LS and HS states. The unique nature of spin crossover materials allows temperature to change the ice rules allowing straightforward access to Coulomb phases not yet observed in water or spin ices. 
%In molecular materials and frameworks frustrated elastic interactions between SCO molecules can give rise to spin-state ices -– phases of matter without long-range order, characterized by a local constraint or ‘ice rule’ and topological excitations that are deconﬁned quasi-particles with spin fractionalized midway between the spins of the LS and HS states. 
\end{abstract}

\maketitle

%\textit{Introduction:}  %Landau theory classifies phases of matter by the symmetries they spontaneously break. This insight has provided deep insights into a huge range of materials \cite{basic}. The topological classification of materials shows that materials can be in different phases even if they have the same symmetry. Much work to date has focused topological materials where the interactions are not important for this. But, there are two well-known examples of topological materials where interactions are crucial: fractional quantum Hall states and (quantum and classical) spin liquids. An important goal is to expand this list. Here we take a small step in that program.
%
%Frustration; the inability to simultaneous minimize over competing interactions can produce macroscopically degenerate states of matter \cite{Frustration} such as spin ices \cite{spinice} and topological quantum spin liquids \cite{Frustration}. These unconventional ground states support exotic excitations, such as mangetic monopoles in spin ices \cite{spinice} and non-abelion anyons in Kitaev quantum spin liquids \cite{Frustration}. Beyond the theoretical interest, macroscopic degeneracy has practical applications as data storage \cite{field_guide}, memory \cite{field_guide}, quantum computation \cite{quantum_computation} and spintronics \cite{1D_SL}.  
%
%Spin ice is a quintessential example of a frustrated magnet. 
In spin ices, such as Dy$_2$Ti$_2$O$_7$ and Ho$_2$Ti$_2$O$_7$, the magnetic Dy (Ho) atoms form a pyrochlore lattice composed of vertex sharing tetrahedra. %The magnetic moments  are constrained to point into or out of each tetrahedra.
The combination of the crystal field and the long-range dipolar interaction constrain the magnetic moments to obey the 2-in/2-out ice rule:
Two of the spins  point into each tetrahedron and two  point out. 
% The ferromagnetic interactions between the spins are suppressed by the dipolar interactions, resulting in a classical macroscopically degenerate state of matter. The degenerate  states satisfy a local constraint or `ice rule' where two of the spins are constrained to point into each tetrahedron and two are constrained to point out. 
The ice rule can be mapped onto a divergence-free flux, analogous to constraints in magnetostatics and electrostatics. Violations of the ice rule carry a fraction of the magnetic spin degreee of freedom, behaving effectively as magnetic monopoles with an emergent Coulombic interaction between them \cite{spinice}. Spin ices are therefore said to be in a  ``Coulomb phase'', which could also arise in frustrated antiferromagnets and some quantum spin liquids  \cite{Henley,Frustration}. 

In principle ices obeying 1-in/3-out or 3-in/1-out ice rules should also give rise to Coulomb phases. This has not yet been observed, although an ordered phase containing 1-in/3-out and 3-in/1-out tetrahedra has been reported \cite{Lefrancois}.
For the nearest neighbor Ising model on the pyrochlore lattice with collinear spins, $S_i$, then, up to a constant, 
\begin{equation}
{\mathcal H}_I=J\sum_{\langle i,j\rangle}S_iS_j+B\sum_iS_i=J\sum_\alpha(L_\alpha+B/4J)^2,
\label{eq:Htoy}
\end{equation}
 where $L_\alpha=\sum_{i\in\alpha}S_i$ and $\alpha$ labels the tetrahedra. The ground state is clearly achieved whenever the magnitude of the `flux', $L_\alpha+B/4J$, is minimized for all $\alpha$. Ice rules require the same value of $L_\alpha$ on all tetrahedra. Therefore, sweeping $B$ moves the ground state between difference ice rules.
However, in spin ices  the spins are not collinear an so one cannot simply apply a magnetic field to change the ice-rules in these systems. 

Recently, it has been proposed that spin-state ices can occur in spin-crossover materials \cite{JaceKagome}. In this Letter we show that  multiple Coulomb  arise for spin-crossover materials on the pyrochlore lattice. Furthermore, we show that sweeping \textit{temperature} alone is sufficient to tune between Coulomb phases obeying three different ice rules.  This is a direct consequence of the competition between the single molecule spin crossover behavior and the many-body ice physics.
We predict that pinch points singularities, the \textit{sine qua non} of the Coulomb phase, will be detectable via neutron scattering in small magnetic fields and that the low-energy excitations are deconfined and carry a spin midway between the that of the two spin states of a single molecule.

Spin crossover (SCO) occurs in transition metal centers in complexes and frameworks when the low-spin (LS; e.g., t$_{2g}^6$e$_g^0$, $S=0$) and high-spin (HS; e.g., t$_{2g}^4$e$_g^2$, $S=2$) states have comparable enthalpy.
SCO provides a reversible molecular switch, which is addressable by changes in temperature, pressure, light irradiation, magnetic field, and chemical environment \cite{gutlich}. As spin-state changes are accompanied by changes in molecular volume, color, and magnetic susceptibility, SCO materials are intrinsically multifunctional and have been widely explored for applications such as  high-density reversible memory, and ultrafast nanoscale switches \cite{gutlich,Kahn,App}. However, many questions about the fundamental physics at play in these systems remain open \cite{Paez-Espejo,StretchBend,Mariette,Pavlik,Nishino,Nishino15,Nishino19,ref1,ref2,ref3,ref4,ref5,Konishi,Stoleriu,JaceSquare}.

\begin{figure} 
	\includegraphics[width=\linewidth]{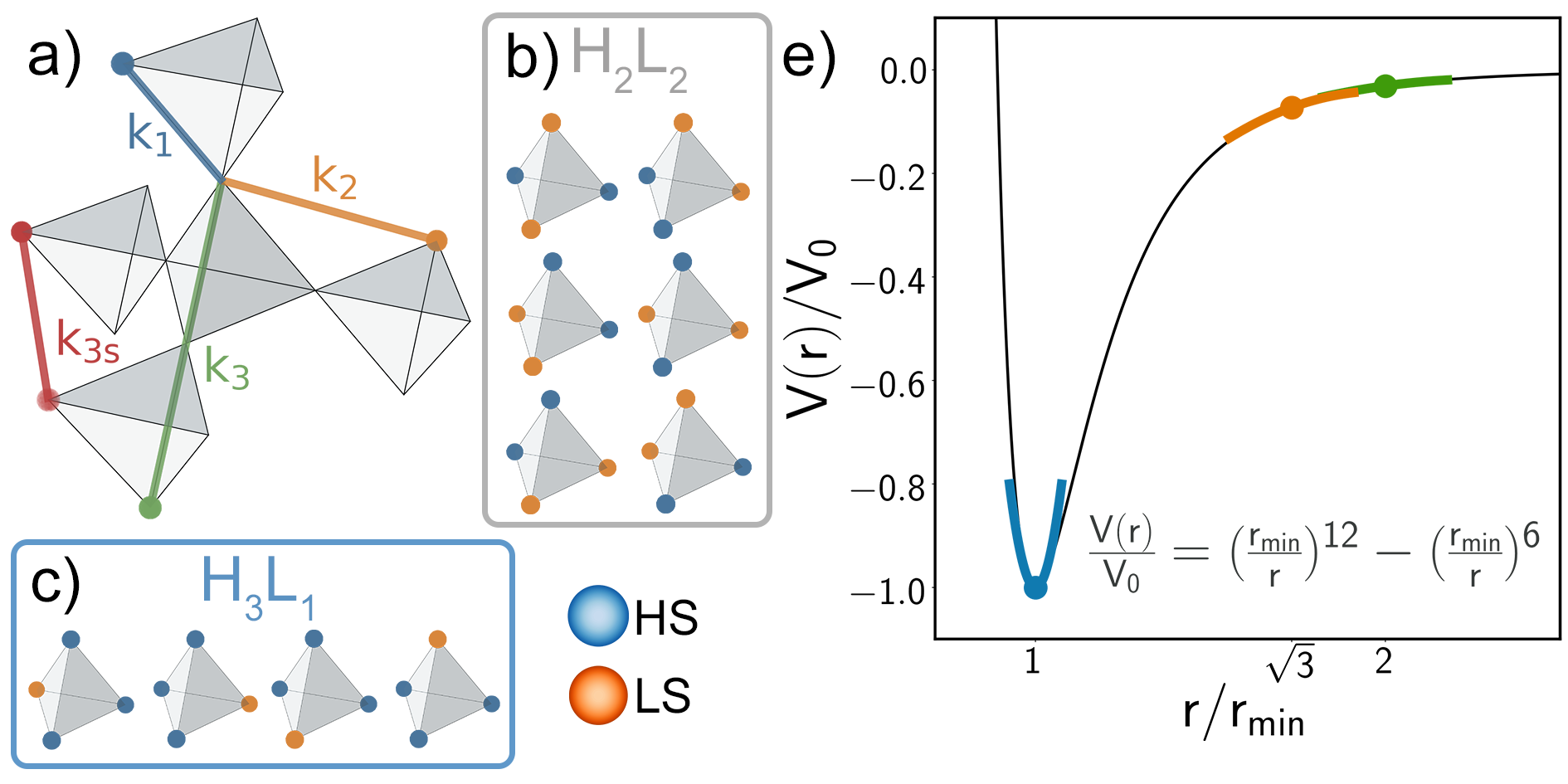}
	\centering
	\caption{
		(a) Pyrochlore lattice with the three nearest neighbor interactions ($k_1$, $k_2$, and $k_3$) marked. We neglect through-space interactions between third nearest neighbors, $k_{3s}$, as we expect these to be much weaker than $k_3$.
		Ice rules in the (b) H$_2$L$_2$ and (c) H$_1$L$_3$ phases. 
		(d) Near the minimum of the potential between neighboring metals  $\partial^2 V(r)/\partial r^2\simeq k_1$, is positive and large. At larger distances the second derivative is negative and decreases in magnitude with increasing distances. Therefore, one expects $k_1>0$ and $k_2<k_3<0$ for the pyrochlore lattice. At larger distances the higher order (non-harmonic) contributions to the potential simply renormalize $\Delta H$ and $k_n$ because $\sigma_i^2=1$. 
	}
	\label{fig:SSIrules_LJP} 
\end{figure}

Changes in the molecular volume accompany spin-state switching due to the (de)population of antibonding $e_g$ orbitals in the (LS) HS state: the metal-ligand bond length in the HS state is often $\sim10~\%$ larger than that in the LS state. In  molecular materials and frameworks, the local structural distortions caused by metals changing spin state couple to long-range elastic interactions. It is convenient to introduce a pseudospin label for the spin state of each metal,  $\sigma_i=1$ ($-1$) if the $i$th molecule is HS (LS). Because  $\sigma_i^2=1$ no terms above quadratic order appear in pairwise interactions %; therefore, regardless of the form of the potential we can model the interactions as elastic  
\cite{JaceSquare}. Letting $R_H$  ($R_L$) be the equilibrium distance between metals in the HS  (LS) phases, we can write the equilibrium midpoint between neighboring metals as $\overline{R} + \delta(\sigma_i+\sigma_j)$, where, $\overline{R}=(R_H+R_L)/2$ and $\delta=(R_H-R_L)/2$. Hence, 
\begin{equation}
\mathcal{H}=%\frac{1}{2}\sum_i(\Delta H-T\Delta S)\sigma_i\equiv 
\frac{\Delta H}{2}\sum_i \sigma_i+
\sum_{n=1}^m\frac{k_n}{2}\sum_{\langle i,j \rangle_n}\left\{r_{ij}-\eta_n\left[\overline{R}+\delta(\sigma_i+\sigma_j)\right]\right\}^2,
\label{eq:HSOC}
\end{equation}
where $\Delta H=H_H-H_L$ is the enthalpy difference between HS and LS metals and $k_n$ are the effective spring constants between $n$th nearest neighbors (see Fig. \ref{fig:SSIrules_LJP}), the sum  over $\langle i,j \rangle_n$ includes all \textit{n}th nearest-neighbors, $r_{ij}$ is the instantaneous distance between sites $i$ and $j$, and $\eta_n=1, \sqrt{3}, 2, \dots$ is the ratio of distances between the \textit{n}th and 1st nearest-neighbor distances on the undistorted pyrochlore lattice. 
Inspired by recent progress in the synthesis  of tetrahedral iron cages \cite{Cages} we  study this model with $m=3$ on the pyrochlore lattice  (\cref{fig:SSIrules_LJP}a), where we expect $k_2<k_3<0$ (\cref{fig:SSIrules_LJP}d) and require  $k_1+6k_2+4k_3>0$ for structural stability. %There are two distinct third nearest neighbor, $k_3$ interactions; the $k_{3t}$ interaction that spans two tetrahedron and the $k_{3h}$ interaction that spans a hexagonal pore. We include the $k_{3t}$ interaction but, not the $k_{3h}$ interaction which one might expect to be much weaker, see Fig \ref{fig:SSIrules_LJP}.

There are two contributions to the entropy. The usual many-body entropy described by the configuration of the $\sigma_i$ and a single molecule term, $\Delta S$, due to changes in the spin and orbital quantum numbers and vibrational frequencies between the HS and LS states. Below we set $\Delta S=4\ln5$, a typical value of Fe(II) complexes  \cite{Entropy}. It is convenient to define $\Delta G=\Delta H-T\Delta S$ and absorb this term into the Hamiltonian  \cite{WP}.
We make the  `symmetric breathing mode approximation', which neglects asymmetric structural distortions  \cite{JaceKagome,JaceSquare}, and calculate the finite temperature properties from a combination of single spin-flip Monte Carlo, worm and loop algorithms and parallel tempering on a $12\times12\times12\times4$ lattice expect where stated.

The weak magnetic interactions in most SCO materials means that the fraction of HS molecules, $n_{HS}\sim\chi T$. Hence the magnetic susceptibility,  $\chi$, is commonly used as a diagnostic for the cooperative behavior in the system. Strong cooperative behaviors often give rise to multiple step transitions \cite{gutlich,Multi1,Multi2,Multi3,Multi4}. Typically, when $k\delta^2\sim k_BT_{1/2}$ \cite{JaceSquare}, where $T_{1/2}=\Delta H/\Delta S$, the temperature at which one expects $n_{HS}=1/2$, is typically 100-400~K. The intermediate plateaus often display long-range patterns of HS and LS metals. But, several disordered phases have been reported also \cite{DisorderKagome,DisorderMat,DisorderOrt}.

\begin{figure} 
	\includegraphics[width=\linewidth]{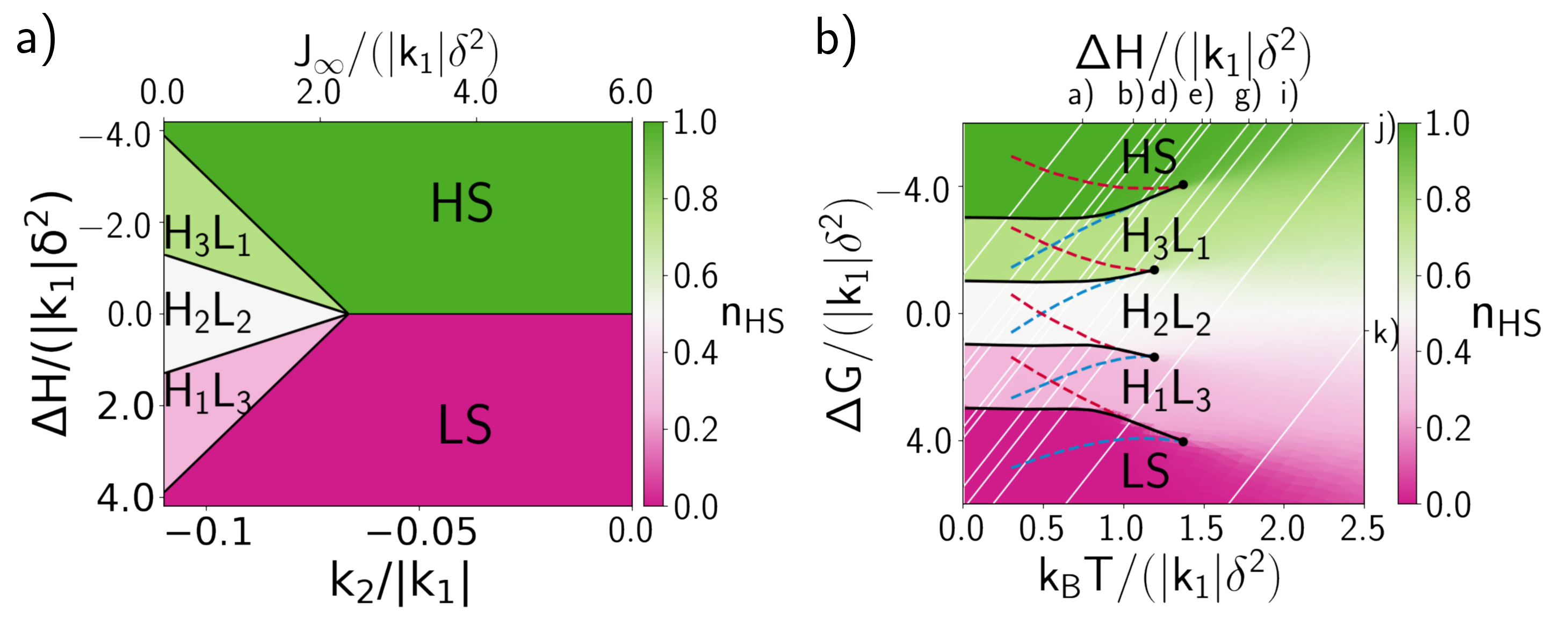}
	\centering
	\caption{
		a) The zero phase diagram for $k_1>0$ and $k_3=3k_2/4$. For small $|k_2|$ the long-range strain dominates and picks out a ferroelastic state, but large $|k_2|$ suppresses this effect allowing the spin-state ices to emerge.
		b) The finite temperature phase diagram for $k_1>0$, $k_2=-0.1k_1$, and $k_3=0.075k_1$. The colors of the phase diagram indicate the equilibrium values of the fraction of high spins $n_{HS}$, calculated via parallel tempering. We find four (black) lines of first order transitions that terminate at critical end points (black circles). The blue and red lines indicate lines of metastability for the cooling and heating calculations, respectively, cf. Fig. \ref{fig:HeatCool} \cite{sup}. Hence, the distance between blue and red lines is the width of the hysteresis. %We find that multiple SSI phases: H$_3$L$_1$ , H$_2$L$_2$ and H$_1$L$_3$, exist over extended regions at finite temperature. 
		Individual materials have a fixed $\Delta H$  (white lines), the corresponding HS fractions, $n_{HS}$, and heat capacities are shown in Fig. \ref{fig:thermo} and Fig. \ref{fig:Cv} \cite{sup}, respectively. 
	}
	\label{fig:ZPT} 
	\label{fig:FPD} 
\end{figure}

We find three distinct spin-state ice (SSI) phases: $\text{H}_3\text{L}_1$, $\text{H}_2\text{L}_2$ and $\text{H}_1\text{L}_3$, see \cref{fig:ZPT} \cite{sup}. In the ground state of the $\text{H}_n\text{L}_{4-n}$ phase every tetrahedron contains $n$ HS and $4-n$ LS metals (Fig. \ref{fig:SSIrules_LJP}b-c). Thus, the $\text{H}_3\text{L}_1$ and $\text{H}_1\text{L}_3$ phases can be mapped onto the dimer model on a diamond lattice, and the $\text{H}_2\text{L}_2$ phase can be mapped onto the loop model on a diamond lattice \cite{Henley}. 

% The corresponding zero-temperature phase diagram, see Fig. \ref{fig:ZPT}, is fully characterized by the two dimensionless ratios: $\Delta H/(\delta^2 k_1)$ - the competition between the tendency of the individual metal centers to undergo SCO and the elastic interactions and $k_2/k_1$ - the relative importance of the next nearest and nearest neighbor elastic interactions. However, it is useful to consider the relative strength of the long-range interactions, $J_\infty=6(k_1+9k_2)$, compared to the elastic interactions. When the long-range strain dominates ($k_1>0$ and $k_2>0$) the ground state is either HS or LS depending on the sign of $\Delta H$. Analogously, when the single molecular enthalpy dominates over the elastic interactions, the ground state is either HS or LS. However, when the elastic interactions dominate over both the long-range strain and enthalpy of the single molecules ($k_1>0$, $k_2<0$ and $\abs{\Delta H} <= -6(k_1+15k_2)\delta^2$), the SSI phases: H$_3$L$_1$, H$_2$L$_2$ and H$_1$L$_3$ are stabilized, see Fig. \ref{fig:SSIrules_LJP}b-c for the corresponding ice rules. 

%\begin{figure} 
%	\includegraphics[width=0.9\linewidth]{Fig4.png}
%	\centering
%	\caption{
%		The finite temperature phase diagram for $k_1>0$ and $k_2=-0.1k_1$. The colours of the phase diagram indicate the equilibrium values of the fraction of high spins $n_{HS}\sim \chi T$, where $\chi$ is the susceptibility, calculated via parallel tempering. We find four (black) lines of first order transitions that terminate at critical end points (black circles). The blue and red lines indicate lines of metastability for the cooling and heating calculations, respectively. Hence, the distance between blue and red lines is the width of the hysteresis. We find that multiple SSI phases: H$_3$L$_1$ , H$_2$L$_2$ and H$_1$L$_3$, exist over extended regions at finite temperature. Individual materials (represented by a white line) have a fixed $\Delta H$. The $n_{HS}$ and $c_V$ for individual materials, labeled a)-l), is shown in Fig. \ref{fig:thermo} and Fig. \ref{fig:Cv}, respectively. Calculations on heating and cooling are shown in Fig. \ref{fig:HeatCool}.
%	}
%	\label{fig:FPD} 
%\end{figure} 

\begin{figure} 
	\includegraphics[width=\linewidth]{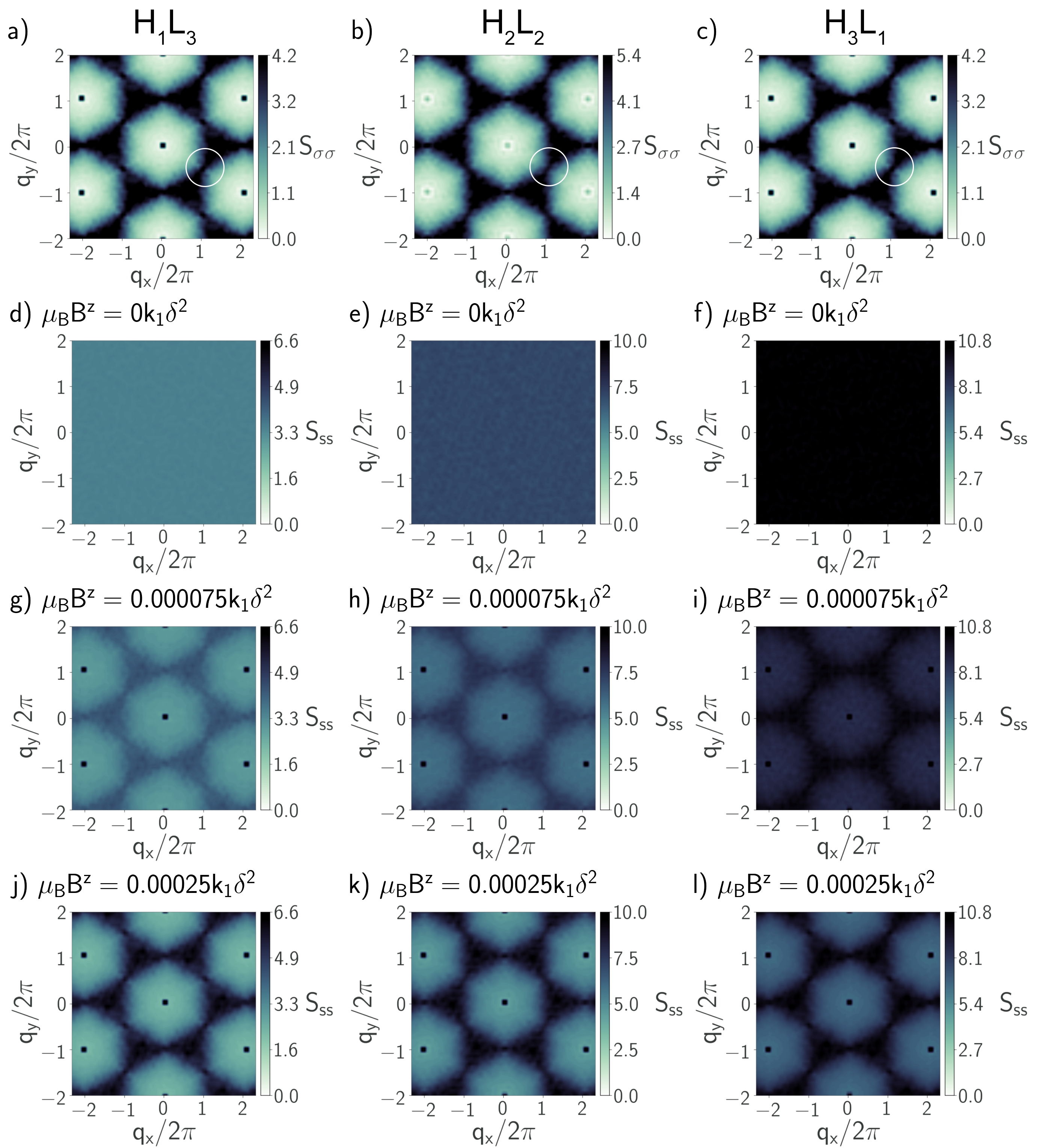}
	\centering
	\caption{
		(a-c) The pseudospin structure factors, $S_{\sigma\sigma}$, Eq. \ref{eq:SF}, for (a) H$_1$L$_3$, (b) H$_2$L$_2$, and (c) H$_3$L$_1$. All structure factors display pinch points, one of which is circled, at the Brillouin zone boundaries characteristic of a Coulomb phase. The existence of the pinch points demonstrates that the low-energy physics of Coulomb phases are described by a divergence-less gauge field \cite{Henley}.
		(d-l) The spin structure factors, $S_{SS}$, Eq. \ref{eq:spinSF}, for (d,g,j) H$_1$L$_3$, (e,h,k) H$_2$L$_2$, and (f,i,l) H$_3$L$_1$ at selected magnetic field strengths. For $k\delta^2\sim100$~K the calculations are $<0.1$~T, so only moderate fields are required for the pinch points to be observable via neutron scattering. 
		 All structure factors were calculated for $k_1>0$, $k_2/k_1=4k_3/3k_1=-0.1$ and $k_BT/(k_1\delta^2)=0.01$ on a $36\times36\times4\times4$ lattice with (a,d,g,j) $\Delta H=2k_1\delta^2$, (b,e,h,k) $\Delta H=0$ and (c,f,i,l) $\Delta H=-2k_1\delta^2$. 
	}
	\label{fig:SF} 
\end{figure}

In order to verify the phases at $T=0$ are indeed Coulomb phases we have calculated the pseudo-spin structure factor, Fig. \ref{fig:SF}j-l,
\begin{equation}
S_{\sigma\sigma}(\vec{q})=\frac{1}{N}\sum_{ij}
%(
\langle\sigma_i\sigma_j\rangle
%-\langle\sigma_i\rangle\langle\sigma_j\rangle) 
e^{i\vec{q}\cdot\vec{r_{ij}}}.\label{eq:SF}
\end{equation}
We clearly observe singularities in $S_{\sigma\sigma}$ at the Brillouin zone boundary, known as pinch points, which are a direct consequence of the existence of a divergenceless gauge field \cite{Henley}. This confirms that the intermediate plateaus are indeed Coulomb phases.

However, directly measuring the pseudospin structure factor is not straightforward. In SCO materials the magnetic correlations between sites are typically negligible. Therefore, the spins are described by a Zeeman Hamiltonian, ${\cal H}_Z=\sum_i\mu_B^z  B \bm{S}_i^z(\sigma_i)$, where the spin, $\bm S_i(\sigma_i)$, depends on the spin-state, $B$ is the applied field, and $\mu_B$ is Bohr magneton. The spin structure factor is 
%Furthermore, the pseudo-spin degrees of freedom are amendable to spin degrees of freedom, allowing for the pseudo-spin structure factor to be measured by neutron scattering experiments. 
%When zero magnetic field is applied the spins associated with the metal centers, $S^z_i$, have $S(S+1)$ possible values. However, applying a large field along the $z$-direction polarizes the metal centers and to a good approximation we can write $S^z_i=\bar{S}+\delta S\sigma_i$. Where $\bar{S}=\frac{1}{2}(S_{HS}+S_{LS})$ and $\delta S=\frac{1}{2}(S_{HS}-S_{LS})$. Consequently, the neutron scattering structure factor can be written as 
\begin{subequations}
\begin{eqnarray}
S_{SS}(\vec{q}) &\equiv& \frac{1}{N}\sum_{ij} \langle \bm S_i(\sigma_i) \cdot \bm S_j(\sigma_j)\rangle  e^{i\vec{q}\cdot\vec{r_{ij}}} \\
%&=& \left[
%	%\langle S_{H}^2\rangle + 
%	\langle S_{L}^2\rangle 
%	%-\langle S^z_i(1) \rangle^2 
%	- \langle S^z_i(-1) \rangle^2\rangle
%\right] 
%\notag\\
%&&+ \left[
%\langle S_{H}^2\rangle - \langle S_{L}^2\rangle 
%-\langle S^z_i(1) \rangle^2 + \langle S^z_i(-1) \rangle^2\rangle
%\right] n_{HS}
%\notag\\
%&&
%	+\left[
%		\frac{1}{4}\left(
%			3m_-^2
%			-m_+^2
%			+2m_+m_-
%		\right)
%%	\right.
%%\notag\\
%%&&
%%	\left.
%		+\left(
%			m_+^2 - m_-^2 
%		\right)
%		n_{HS}
%	\right]\delta(q)
%\notag\\
%&&+\left(\frac{m_+^2 + m_-^2 - 2m_+m_-}{4} \right)S_{\sigma\sigma}(\vec{q})\\
&=&
	\left[
		m_+^2 + m_-^2 - 2m_+m_- 
	\right]
	\frac{S_{\sigma\sigma}(\vec{q})}{4} 
%	\notag\\
%	&=&
%	\left[
%	\langle S^z(1) \rangle^2+\langle S^z(-1) \rangle^2-2\langle S^z(-1) \rangle\langle S^z(1)\rangle 
%	\right]
%	\frac{S_{\sigma\sigma}(\vec{q})}{4}
	\notag\\
&&
+S_d(B^z)+S_B(B^z)\delta(q),
\end{eqnarray}
\label{eq:spinSF}
\end{subequations}
where $m_\pm=\sum_i\langle S_i^z(\pm1)\rangle/N$, and expressions for the trivial diffuse, $S_d(B^z)$ and Bragg, $S_B(B^z)$, scattering are given in \cite{sup}.
This, is directly measurable via neutron scattering and clearly shows the pinch points,  Fig. \ref{fig:SF}. We note that the fields required are modest. For $k\delta^2\sim100$~K the calculations are $<0.1$~T.
%\textcolor{red}{update to reflect new calcs} 
%\textcolor{blue}{Jace - The problem with using the new cals is that the mapping between the two terms becomes quite messy. $S_{zz}=\overline{S}(\overline{S}+2\delta Sm)\delta(\vec{q})+\delta S^2S_{\sigma\sigma}$, where $m=\langle \sigma_i \rangle$} 
%\textcolor{red}{Ben - still need to explain what we're plotting - it's a prl - no one is meant to understand it without putting in some hard yards.}

\begin{figure} 
	\includegraphics[width=\linewidth]{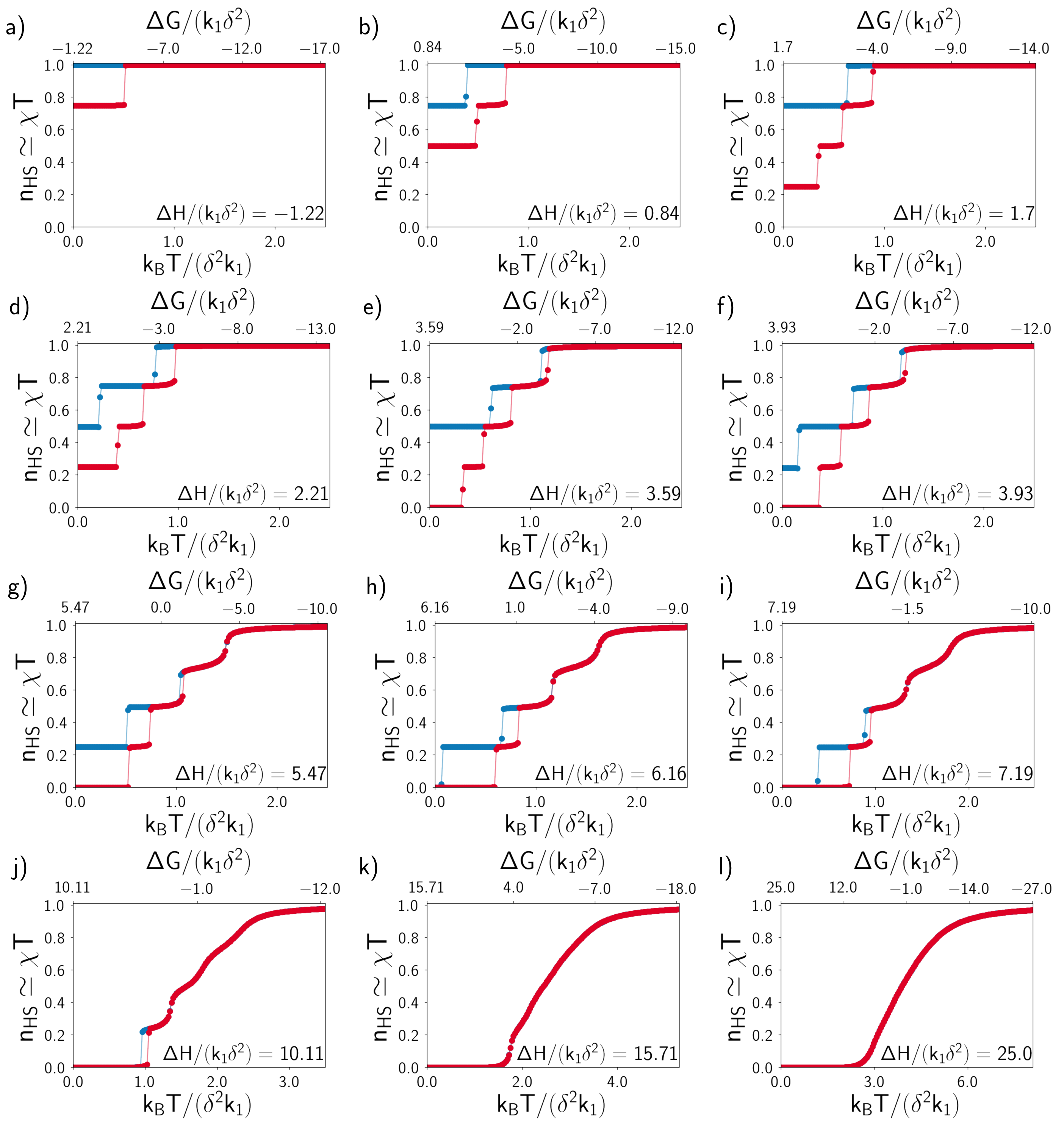}
	\centering
	\caption{
		The fraction of high spins, $n_{HS}$, for selected values of $\Delta H/(k_1\delta^2)$ with $k_1>0$ and $k_2/k_1=4k_3/3k_1=-0.1$. The red and blue lines represent heating and cooling respectively. 
		%$\Delta G=\Delta H-T\Delta S$ is marked not the upper $x$-axis of each plot; the value at $T=0$ equals $\Delta H$. 
		We find plateaus at $n_{HS}\simeq0$, $\frac{1}{4}$, $\frac{1}{2}$, $\frac{3}{4}$ and $1$, corresponding to the LS, H$_1$L$_3$, H$_2$L$_2$, H$_3$L$_1$, and HS phases. Hence, for a wide range of parameters,  it is possible to tune between different Coulomb phases with temperature alone. %Snapshots of these phases are shown in Fig. \ref{fig:defects}.
	}
	\label{fig:thermo} 
\end{figure}

%We find SSI phases not just at zero-temperature but, extended regions above $T=0$. For a fixed $k_1>0$, $k_2<-1/15$ and $J_\infty>0$ we observe $10$ different types of thermodynamic behaviours and four-critical points in the finite temperature phase diagram. A typical slice of the the finite temperature phase diagram for this parameter regime is shown in Fig. \ref{fig:FPD}. 
The ratio $\Delta H/(\delta^2 k_1)$ not only has important consequences for the low-temperature physics, but also for the high-temperature behavior as well, Figs. \ref{fig:FPD}b and  \ref{fig:thermo}. For   $\Delta H \sim -(k_1\delta^2)$ we observe a single first order transition. This is a purely collective phenomena, as the single ion physics always favors the HS state ($\Delta G<0\,\forall\, T$). Increasing $\Delta H$ induces further transitions with plateaus at $n_{HS}\simeq0$, $\frac{1}{4}$, $\frac{1}{2}$, $\frac{3}{4}$ and $1$, corresponding to the LS, H$_1$L$_3$, H$_2$L$_2$, H$_3$L$_1$ and HS phases respectively, \ref{fig:thermo}b-h. Hence, for a wide range of parameters,  it is possible to tune between different Coulomb phases with temperature alone. 

This can be understood as follows: The single molecule entropy difference between spin-states ($\Delta S$) can be absorbed into the Hamiltonian \cite{WP}. Thus, $\Delta G=\Delta H-T\Delta S$ replaces $\Delta H$ in \cref{eq:HSOC}. This term couples to the pseudospin just as a magnetic field couples to spin in the Ising model, \cref{eq:Htoy}. That is, the single molecule spin crossover behavior acts as an effective temperature-dependent `field' for the pseudospins. This changes the ice rules as the temperature varies.

%It should be noted that due to the intrinsic difference in entropy of HS and LS states, as $T\rightarrow\infty$ the high-spin fraction, for Fe(II) metal centers $\Delta S=4k_B\log(5)$, $n_{HS}\rightarrow e^{\Delta S/k_B}/(1+e^{\Delta S/k_B})=625/626\approx 0.9984$. These plots have stable phases consistent with plateaus at $n_{HS}=0$ (LS), $\frac{1}{4}$ (H$_1$L$_3$), $\frac{1}{2}$ (H$_2$L$_2$), $\frac{3}{4}$ (H$_3$L$_1$) and $1$ (HS). Hence, it is possible to tune between different SSI phases with temperature alone. Snapshots of these phases are shown in Fig. \ref{fig:defects}. 

Due to the large width of the hysteresis loop at low temperatures, simulations of straightforward cooling  does not always result in the same low temperature phase as is found by parallel tempering, Fig. \ref{fig:thermo}a-g. Similar effects have been observed experimentally in SCO materials that display long-range antiferroelastic order \cite{Milin}, and labeled ``hidden hysteresis''. The hidden low temperature states  can be realized by either  photoswitching  (i.e., reverse-light induced excited spin-state trapping) or  applying and adiabatically releasing a pressure to  the system. Hence, it becomes possible to tune between different SSI phases with not only temperature but, pressure and light as well. 

\begin{figure} 
	\includegraphics[width=\linewidth]{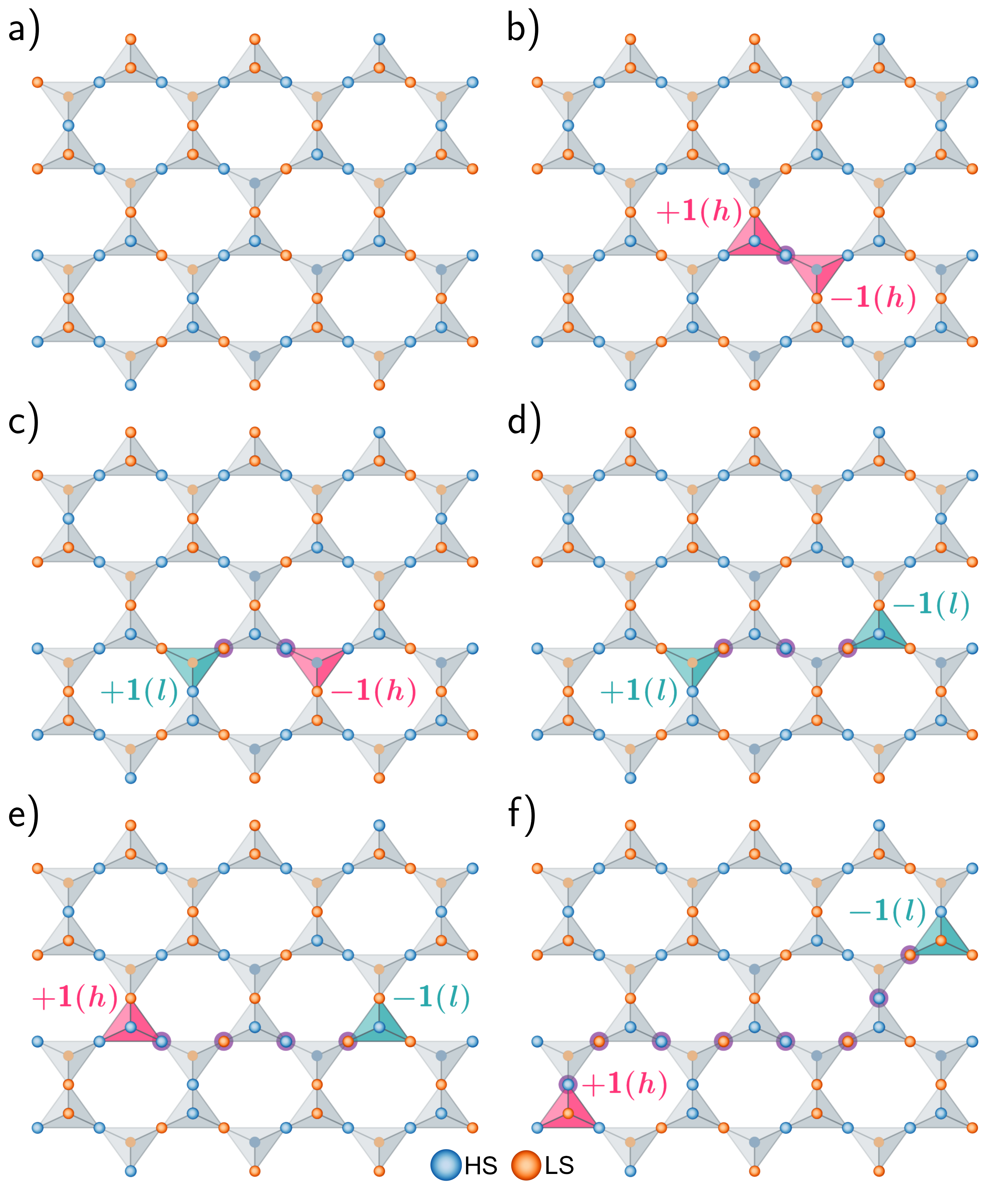}
	\centering
	\caption{
		Sketch of the propagation of defects in the H$_2$L$_2$ phase. (a) The H$_2$L$_2$ vacuum state,  every tetrahedron obeys the 2HS/2LS spin-state ice rule. (b) Changing the spin-state of a LS metal center (highlighted in purple) creates  $h$ defects, on both tetrahedra connected to the metal, but does not change the total topological charge, $\sum Q=\sum\kappa \delta S$, where, $\kappa=\pm 1$ for tetrahedra pointing in the $\pm z$ direction, and the sum runs over all tetrahedra. 
		(c-f) Additional spin-state changes on tetrahedra hosting defects are low-energy process and cause the defects to propagate. This conserves $Q$, but not the numbers of $h$ or $\ell$ defects. Similar processes result in deconfined excitations in both the H$_1$L$_3$ and H$_3$L$_1$ SSI phases.
	}
	\label{fig:propagate_defects} 
\end{figure}

Further increasing $\Delta H/(k_1\delta^2)$ moves the transitions towards and through critical points, where the transition is continuous, and into the crossover regime, Figs. \ref{fig:FPD}b and \ref{fig:thermo}g-i. The higher temperature transitions become crossovers first as $\Delta H/(k_1\delta^2)$ increases until there is eventual only a single crossover.
This  results in significant melting of the SSI phases and the spontaneous production of defects. For each SSI phase there are two different types of defects. For a state obeying the spin-state ice rules everywhere changing a metal from a LS to HS state creates a $h$-defect on both of the tetrahedra connected to the metal center, \cref{fig:propagate_defects}a,b. Conversely, changing the spin-state of a metal from HS to LS creates two $\ell$-defects. 
%The $h$ and $\ell$ defects for each of the SSI phases are shown in Fig. \ref{fig:defects}.

To understand these defects it is helpful to consider a large magnetic field in the $z$ direction, such that $S_{i}^z(1)=S_{H}$ and $S_{i}^z(-1)=S_{L}$. 
%For a state that obeys the spin-state ice rules everywhere the corresponding vacuum has the spin %$S_{vac}^z=\frac{1}{2}\sum_{i\in \boxtimes}S^z_i$. Where the sum runs over the metal centers sitting on the corners of a tetrahedron obeying the spin-state ice rules, the factor of a half accounts for the fact that each metal sits on the corners of two tetrahedron. 
%$S_{vac}^z=\sum_{i}S^z_i$.
The creation of two $h$ defects increase the number of HS metals by one. Thus, each defect carries a spin $\frac{1}{2}(S_{H}-S_{L})\equiv\delta S$. Similarly, the process of creating two $\ell$ defects on the connected tetrahedra corresponds to the creation of two quasi-particles with spin $-\delta S$. %Therefore, the $h$ and $\ell$ defects carry only half the spin associated with a single metal changing spin-state. 
It is important to note that there are no intermediate spin-states in the model. Hence, these defects arise purely as a collective effect and thus correspond to fractionalized quasi-particles with spin midway between HS and LS states. 

The multiple of ways to satisfy the spin-state ice rules allow defects to propagate, Fig. \ref{fig:propagate_defects}. 
For example, a metal center  changing from HS to LS on a tetrahedron containing a $h$ defect  restores the ice rules on that tetrahedron and creates a $\ell$ defect on the other tetrahedron connected to the metal center, \cref{fig:propagate_defects}b,c. %This process of propagating defects interconverts $h$ and $\ell$ type defects and hence, 
Thus, the number of $h$ and $\ell$ type defects are not conserved. However, the topological charge $Q=\kappa \delta S$ is conserved, where $\kappa=\pm 1$ for tetrahedra pointing in the $\pm z$ direction. It only takes a finite amount of energy to move a pair of defects infinitely far apart. Hence, the fractionalized topological charges are deconfined \cite{Henley}. 

$h$ and $\ell$ defects carry the opposite spin. Therefore, if  $\delta S$ and $-\delta S$ topological charges meet on the same tetrahedron  they annihilate, restoring the spin-state ice rules.

It has recently been show that strain can induce the motion of domain walls in ordered phases of SCO materials \cite{Morgan}. This is likely a consequence of the volume difference in the HS and LS states. Suggesting that strain should also induce motion of the defects in SSI Coulomb phases. Alternatively, spin-orbit coupling should couple the quasiparticles to an applied voltage. Either of these effects could make SSI a valuable resource for spintronic applications.

%\textit{Conclusion}
Our calculations predict that three distinct Coulomb phases arise in pryochlore lattices in extended regions of both the zero-temperature and finite temperature phase diagrams. In each  phase the low energy excitations are mobile and deconfined, carrying a spin midway between that of the HS and LS states. Realizing Coulomb phases beyond the 2-in/2-out phases in water and spin ices has proven extremely challenging. However, the unique role of the single molecule entropy in spin crossover materials allows  temperature to change the ice rules. The physics of SCO molecules could also allow for the use of pressure and light to manipulate and control the excitations.
  %This opens up the possibility for elastically frustrated pryochlore lattice materials to be used study the spin-Seebeck effect. Thus, opening up the possibility of SCO materials to be used as thermoelectic materials.  
 
Important questions arising from this work include: what are the leading quantum mechanical corrections to the Hamiltonian? and what state do they lead to? Based on analogies to spin ice and the dimer/loop models on the diamond lattice \cite{Frustration,spinice} one would expect that significant tunneling between the classical ground states would replace the multiple SSI phases with $U(1)$ quantum spin-state liquid phases. %The existence of $U(1)$ gauge field would result in artificial light in SCO materials \cite{Frustration}.

\begin{acknowledgments}
This work was funded by the Australian Research Council through grant number DP200100305 and an Australian Government Research Training Program Scholarship.
\end{acknowledgments}

\pagebreak
\clearpage

\begin{widetext}

\renewcommand\thefigure{S\arabic{figure}}
\renewcommand\thesection{S\arabic{section}}
\renewcommand\theequation{S\arabic{equation}}
\renewcommand\thesubsection{S\arabic{section}.\arabic{subsection}}
\setcounter{figure}{0} 
\setcounter{equation}{0} 

\begin{Large}
	\begin{center}
		\textbf{Supplementary Information for ``Multiple Coulomb phases with temperature tunable ice rules in pyrochlore spin crossover materials''}\\
		Jace Cruddas and B. J. Powell
	\end{center}
\end{Large}

\begin{eqnarray}
S_d(B^z)
%&=&\frac{S_{H}\left(S_{H}+1\right)+S_{L}\left(S_{L}+1\right) - m_+^2 - m_-^2}{2} 
%\notag\\
%&&
%+\left(\frac{S_{H}\left(S_{H}+1\right)-S_{L}\left(S_{L}+1\right) - m_+^2+m_-^2}{2} \right)(2n_{HS}-1)\\
&=&S_{L}\left(S_{L}+1\right)  - m_-^2 
+\left[S_{H}\left(S_{H}+1\right)-S_{L}\left(S_{L}+1\right) - m_+^2+m_-^2 \right]n_{HS}\\
S_B(B^z)&=&\left(\frac{m_+^2+m_-^2+2m_+m_-}{4} \right)+\left(\frac{m_+^2-m_-^2}{2} \right)\left(2n_{HS}-1\right).
\end{eqnarray}

%\begin{figure*}[b]
%	\includegraphics[scale=0.5]{Fig_neighbours.png}
%	\centering
%	\caption{
%		First (a), second (b) and third (c) nearest neighbours on the pyrochlore lattice. For the third nearest neighbours we only consider the through- bond neighbours and neglect the through space neighbours.
%	}
%	\label{fig:NNs} 
%\end{figure*}

\begin{figure*}[b] 
	\includegraphics[width=0.6\textwidth]{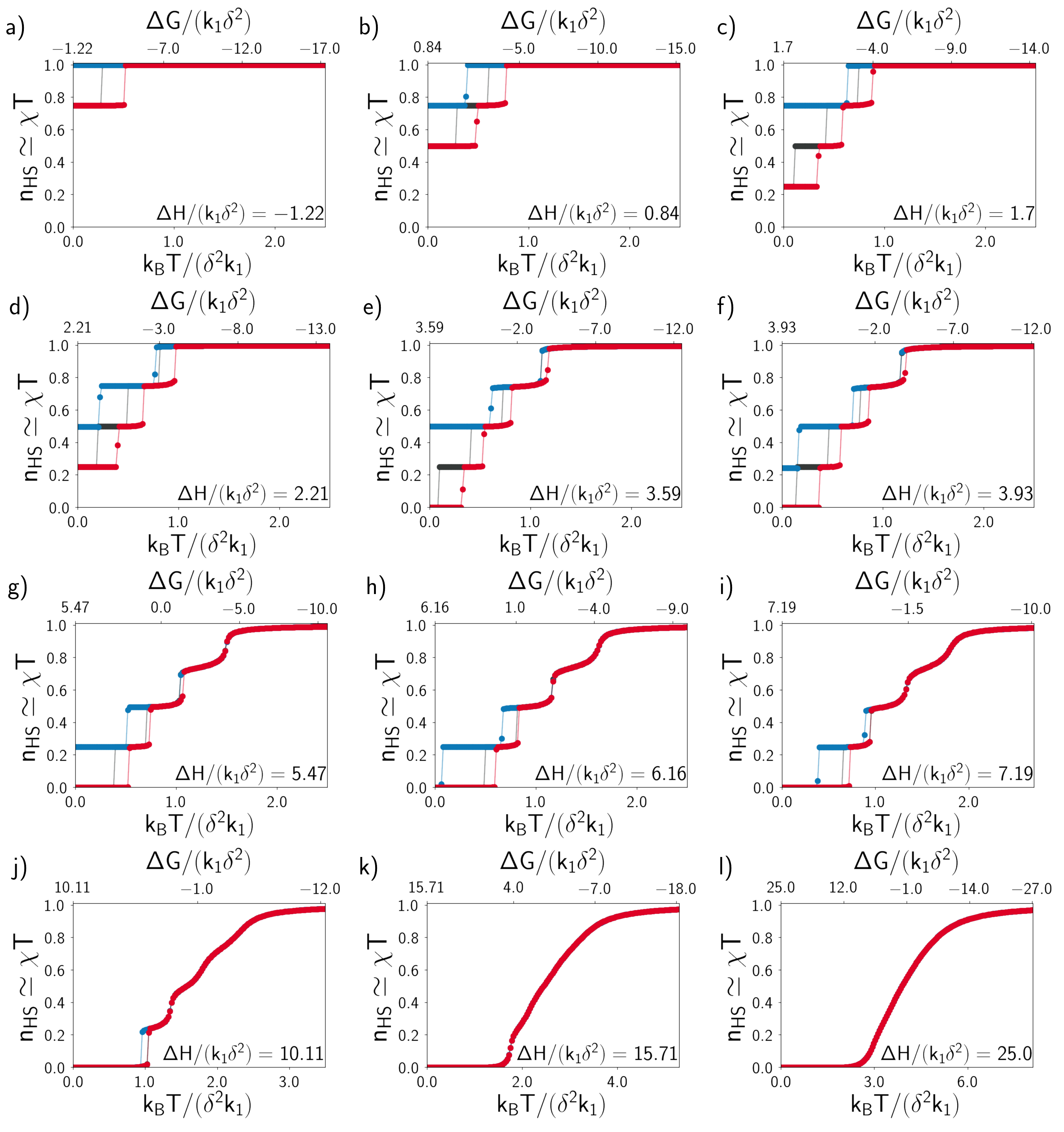}
	\centering
	\caption{
		 (As Fig. \ref{fig:thermo} but with results for parallel tempering added.) The fraction of high spins, $n_{HS}$, for selected values of $\Delta H/(k_1\delta^2)$ with $k_1>0$ and $k_2/k_1=4k_3/3k_1=-0.1$. The red, blue and black lines represent heating, cooling and parallel tempering results, respectively.
%		$\Delta G=\Delta H-T\Delta S$ is marked not the upper $x$-axis of each plot; the value at $T=0$ equals $\Delta H$. 
%		We find plateaus at $n_{HS}\simeq0$, $\frac{1}{4}$, $\frac{1}{2}$, $\frac{3}{4}$ and $1$, corresponding to the LS, H$_1$L$_3$, H$_2$L$_2$, H$_3$L$_1$, and HS phases. Hence, for a wide range of parameters,  it is possible to tune between different Coulomb phases with temperature alone. %Snapshots of these phases are shown in Fig. \ref{fig:defects}.
	}
	\label{fig:thermo3} 
\end{figure*} 

\begin{figure*} 
	\includegraphics[width=0.6\textwidth]{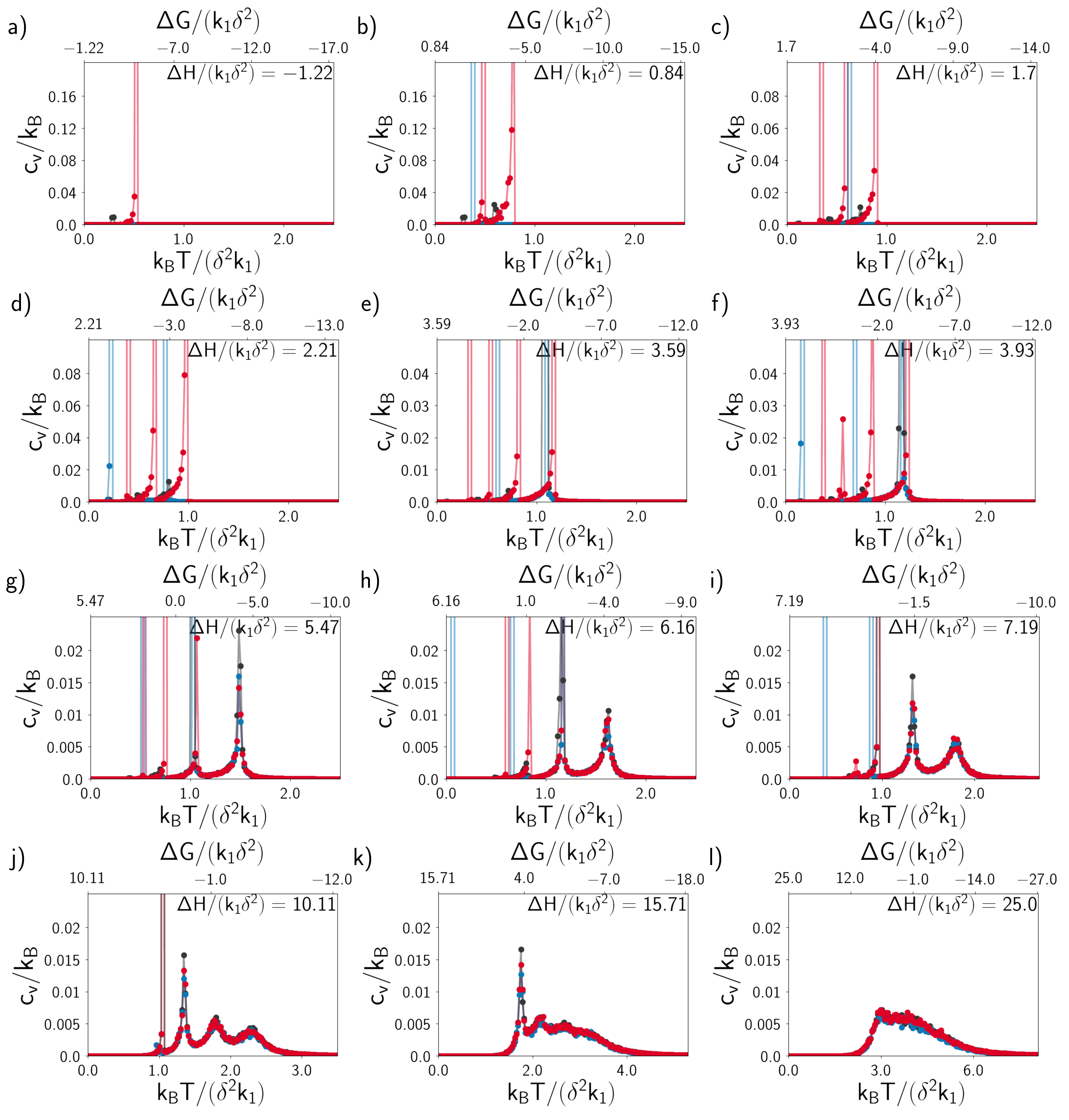}
	\centering
	\caption{
		The heat capacity per site $c_V$ for the same values of $\Delta H/(k_1\delta^2)$ as Fig. \ref{fig:thermo} with fixed values of $k_1>0$ and $k_2=4k_3/3=-0.1k_1$. The black, blue and red lines have the same meaning as Fig. \ref{fig:thermo}. Similarly, for increasing values of $\Delta H/(k_1\delta^2)$ we observe (in the equilibrium values) (a) a first order-transition, (b) two first order transitions, (c-d) three first order transitions, (e-f) four first order transitions, (f-g) three first order transitions and a crossover, (i) two first order transitions and two crossovers, (j) one first order transition and three crossovers, (k) two crossovers and (l) a crossover. Heating and cooling the system would result in (a) no transitions, (b-c) one first order transition, (d-e) two first order transitions, (f) a three-step transition and (g) two-first order transitions. 
	}
	\label{fig:Cv} 
\end{figure*}

\begin{figure*}
	\includegraphics[width=0.7\textwidth]{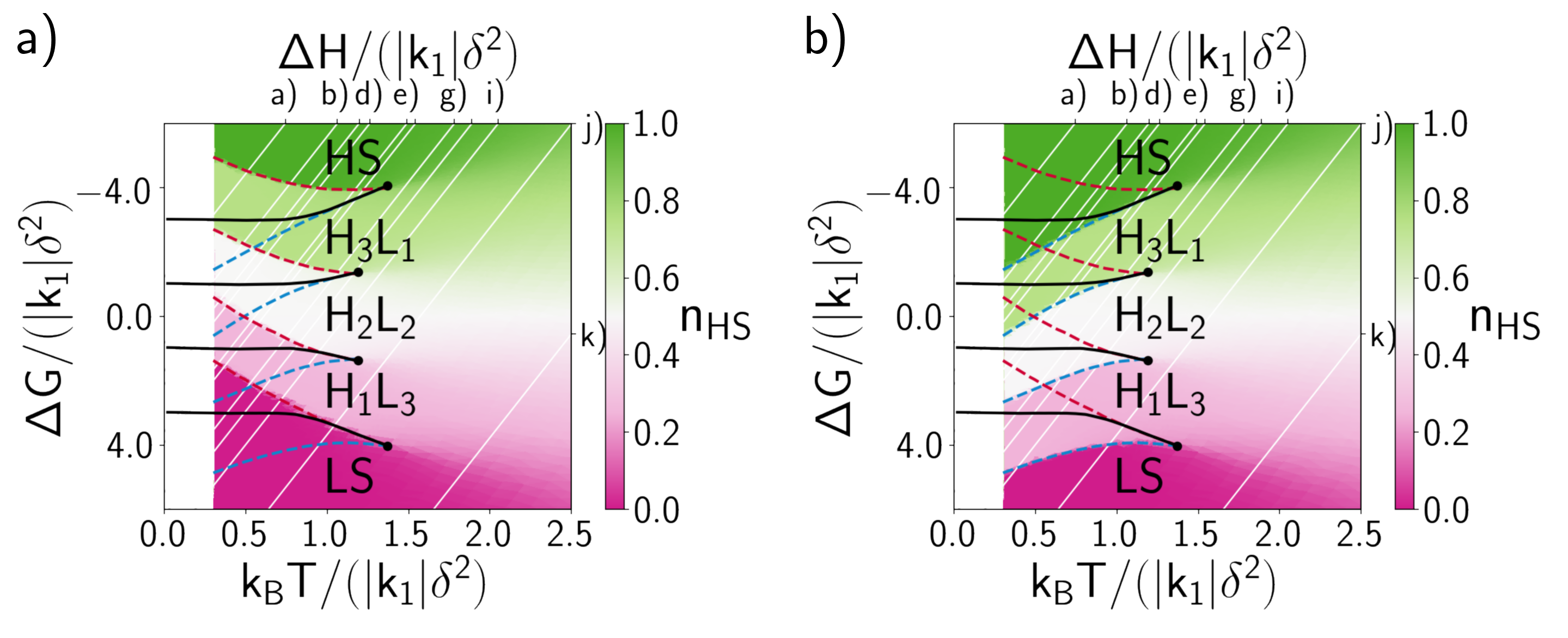} 
	\centering
	\caption{The HS fraction, $n_{HS}$, calculated on (a) cooling and (b) heating for the third nearest neighbor elastic model on the pyrochlore lattice with fixed values of $k_1>0$ and $k_2=(4/3)k_3=-0.1k_1$. Lines and dots have the same meanings as in Fig. \ref{fig:thermo}a.
	}
	\label{fig:HeatCool} 
\end{figure*}

\begin{figure*} 
	\includegraphics[width=\linewidth]{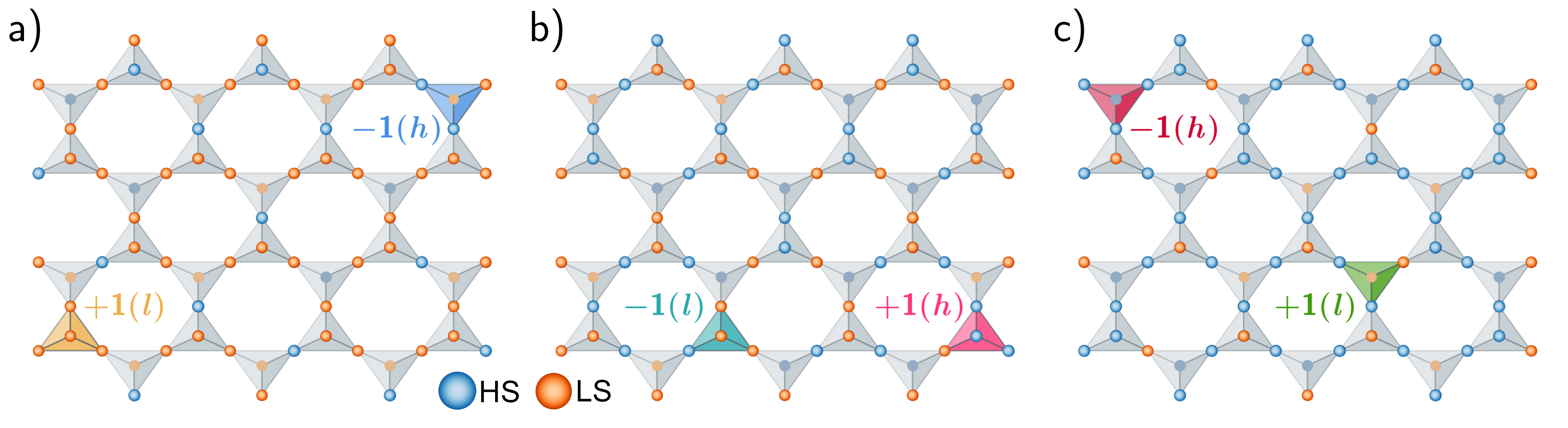}
	\centering
	\caption{ Truncated snapshots of (a) H$_1$L$_3$, (b) H$_2$L$_2$ and (c) H$_3$L$_1$ for $k_1>0$ and $k_2=4k_3/2=-0.1k_1$. Almost all triangles obey the spin-state ice rules, Fig. \ref{fig:SSIrules_LJP}. There are two expectations to the rule; one tetrahedron containing an excess of HS ($h$) and one containing an excess of LS ($\ell$).
		%To understand these defects it is helpful to consider a field applied in the $z$ direction, in which case each $h$ defect carries a total spin $S^z_h-S_{vac}^z=\delta S$, and each $\ell$ defects carries a total spin $S^z_\ell-S_{vac}^z=-\delta S$. Where $\delta S=(S_{HS}-S_{LS})/2$ and $S_{vac}^z$ is the total spin of each tetrahedron in the vacuum state. It is important to note that there are no intermediate spin-state in this model. The fractionalization is purely a collective effect. Due to the multiple ways to satisfy the ice rules, changing a spin-state of a metal center can interconvert a $\ell$ and a $h$ defect (or vice versa) allowing for the propagation of defects, see Fig. \ref{fig:propagate_defects}. 
		The snapshots are taken for (a) $\Delta H=10.11k_1\delta^2$ and $k_BT=1.12k_1\delta^2$, (b) $\Delta H=5.47k_1\delta^2$ and $k_BT=0.76k_1\delta^2$ and (c) $\Delta H=5.47k_1\delta^2$ and $k_BT=1.11k_1\delta^2$.
	}
	\label{fig:defects} 
\end{figure*}

%
%\begin{table}[b]\label{table}
%	\centering
%	\begin{tabular}{|c|c|c|c|c|}
%		\hline 
%		Phase & Spin-state Ice rule & $S^z_{vac}$ & $h$ & $\ell$\\ 
%		\hline 
%		H$_1$L$_3$ & 1 HS; 3 LS & $1/2$ & 2 HS; 2LS & 0 HS; 4 LS  \\ 
%		\hline 
%		H$_2$L$_2$ & 2 HS; 2 LS & $1$ & 3 HS; 1LS & 1 HS; 3 LS \\ 
%		\hline 
%		H$_3$L$_1$ & 3 HS; 1 LS & $3/2$ & 4 HS; 0LS & 2 HS; 2 LS \\ 
%		\hline 
%	\end{tabular} 
%	\caption{Corresponding spin-states rules, and spin of the $h$ and $\ell$ defects for the SSI phases: H$_1$L$_3$, H$_2$L$_2$ and H$_3$L$_1$. The $h$ and $\ell$ defects have spin $S^z_{h}-S^z_{vac}=\delta S$ and $S^z_{\ell}-S^z_{vac}=-\delta S$, respectively.}
%\end{table} 

\end{widetext}

\end{document}